\documentclass[prl,twocolumn]{revtex4}%

\usepackage{amsfonts}
\usepackage{amsmath}
\usepackage{amssymb}
\usepackage{graphicx}

\begin{document}

\preprint{cond-mat/xxx}

\title{Temperature dependence of the magnetic penetration depth in B$_{1-x}$K$_{x}$BiO$_{3}$ superconductor}

\author{Alexey Snezhko} \email{snezhko@physics.sc.edu}%
\affiliation{Department of Physics and Astronomy  and NanoCenter, University of South Carolina, 712 Main St,
Columbia, SC 29208, USA}

\author{Ruslan Prozorov} \email{prozorov@sc.edu}
\affiliation{Department of Physics and Astronomy  and NanoCenter, University of South Carolina, 712 Main St,
Columbia, SC 29208, USA}%

\begin{abstract}
Temperature dependence of the magnetic penetration depth, $\lambda$(T), was measured in the
Meissner state in single crystals B$_{1-x}$K$_{x}$BiO$_{3}$ (x=0.37) using tunnel diode resonator
technique. At low temperatures, $0.013 \le T/T_c  \le 0.4$, $\lambda$(T) is exponentially flat,
which provides a strong evidence for conventional s-wave BCS behavior. Numerical analysis of the
data rules out the possibility of a gap with nodes.
\end{abstract}

\received{25 September 2003}%
\accepted{11 December, 2003}%

\maketitle

Magnetic penetration depth is an effective tool to study electromagnetic properties of
superconductors. At low temperatures, its temperature dependence is directly related to the density
of low energy quasiparticles, which in turn can be related to the anisotropy of the superconducting
energy gap on the Fermi surface. For investigation of the low lying excitations and thus the
anisotropy of the energy gap, the analysis is considerably less ambiguous if measurements are
performed on high quality single crystal and temperatures well below $T_c/3$. In this paper we
report magnetic penetration depth measurements on single crystals B$_{0.63}$K$_{0.37}$BiO$_{3}$
($T_c \approx 31$ K) down to 0.4 K. The investigation of mechanisms of superconductivity in
B$_{1-x}$K$_{x}$BiO$_{3}$ (BKBO) system has been one of the important subjects in studies of high
$T_c$ superconductivity in oxide materials. The significance of BKBO lies in observations that some
of its superconducting properties are consistent with the conventional s-wave isotropic
superconductivity, but others are resembling high-$T_c$ cuprates. In particular, substantial
isotope effect \cite{Loong,Batlogg}, strong phonon contribution from the neutron scattering
measurements \cite{Loong2}, and a superconducting gap with $2\Delta_0/T_c$=3.5$\pm$0.3 from the
tunnelling and optical experiments \cite{Huang,schl,sato,shari} indicate significant role of
electron-phonon interactions in mechanism of superconductivity of BKBO. However the low density of
states at the Fermi level with $T_c$ as high as 30K \cite{Batlogg} and insulator-superconductor
transition by doping \cite{cava} are similar to high-$T_c$ cuprates. However, in contrast to high
$T_c$ cuprates which have two dimensional $CuO_2$ planes, BKBO has simple three dimensional cubic
perovskite structure.

\begin{figure}[tb]
\includegraphics [width=7.5 cm]{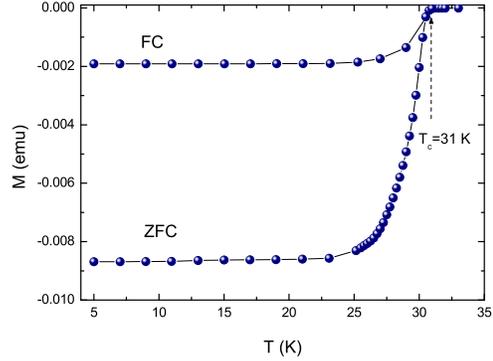}
\caption{Temperature dependence of magnetic moment in B-K-Bi-O single crystal in zero-field cooled and field
cooled experiment in an external applied field of 10 Oe. } \label{fig1}
\end{figure}

Single crystal of  B$_{1-x}$K$_{x}$BiO$_{3}$ (x=0.37) was grown by the electro-chemical method
reported elsewhere \cite{Kim2003}. DC magnetization was measured by using \emph{Quantum Design}
MPMS SQUID magnetometer. Zero-field cooled and field cooled temperature scans taken in external
field of 10 Oe are shown in Fig.~\ref{fig1}. Superconducting transition temperature of the sample
is $T_c\approx31$K and the curves show regular superconducting screening with the Meissner
expulsion of about 20\%, which provides an indication of relatively low pinning. Magnetization
loops confirm low pinning behavior.

\begin{figure}[tb]
\includegraphics[width=7.5cm]{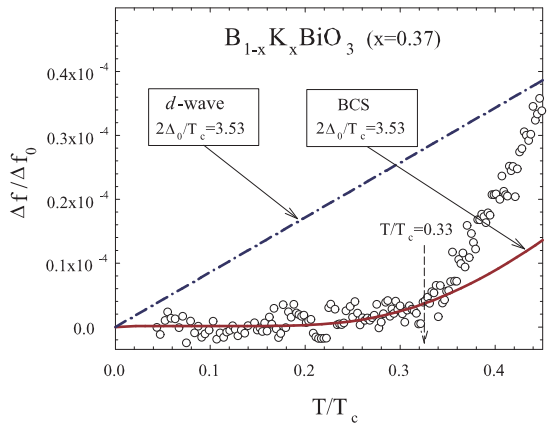}
\caption{Low temperature penetration depth variation in B$_{1-x}$K$_{x}$BiO$_{3}$ (x=0.37) single crystal.
The solid line shows the low temperature exponential fit to the weak-coupling BCS expression. Dash-dotted
line represents low temperature \emph{d}-wave behavior. See text for details.} \label{fig2}
\end{figure}

Penetration depth measurements were carried out using a 13 MHz tunnel diode LC resonator
\cite{carrington,prozorov2} mounted on a He$^{3}$ cryostat. The sample was placed on the sapphire
stage with temperature control from 0.35 to 40K. During the measurements the sample is exposed to
the small \emph{ac} field H$_{ac}\simeq 20$ mOe much less than the first critical field, $H_{c1}
\approx 750$ Oe at 5 K \cite{kimht}. The relative resonance frequency shift, $\Delta f=
f(T)-f(T_{min})$, is related to the change of the magnetic penetration depth via $\Delta f=-G
\Delta\lambda$, where $G$ is geometrical factor that depends upon the sample shape and volume as
well as the coil geometry \cite{prozorov2}. Low temperature behavior of magnetic penetration depth
is shown in the Fig.\ref{fig2}. The frequency shift in Fig. \ref{fig2} was normalized by the value
$\Delta f_{0} \approx 3500$ Hz, which represents the total frequency shift cooling from normal to
superconducting state. The solid line in Fig. \ref{fig2} shows the fit to a low temperature BCS
expression for an s-wave material,

\begin{equation}  \label{2}
\Delta \lambda \approx\lambda (0)\sqrt {\frac{\pi \Delta _0 }{2T}}
\exp \left( {-\frac{\Delta _0 }{T}} \right)
\end{equation}

Here $\Delta_{0}$ is the value of the energy gap at zero temperature \cite{annet}. The fitting
range was chosen up to 0.33$T_c$ to ensure the validity of the low temperature expansion.
$2\Delta_0/T_c=3.53$ corresponds to standard weak coupling BCS value. The dash-dotted line shows
the low temperature behavior of the magnetic penetration depth predicted for a clean d-wave
superconductor \cite{hilsh}. The value of $2\Delta_0/T_c$ for d-wave case was again chosen to be
3.53 in accordance with results of tunnelling and optical experiments.

\begin{equation}  \label{3}
\Delta \lambda \approx\lambda (0) \left( 1-\frac{\ln2}{\Delta _0}T
\right)
\end{equation}

Clearly the isotropic s-wave BCS curve provides best description of the low temperature penetration
depth variation indicating the isotropic nature of superconducting gap for BKBO. Some apparent
noise in the data is because in the temperature interval of interest, the penetration depth is
exponentially flat with no systematic temperature dependence.

\begin{flushleft}
\textbf{Acknowledgements} The authors thank Dr. Hyun-Tak Kim at ETRI for sending us BKBO crystals
grown by Dr. W. Schmidbauer and Prof. J. W. Hodby at University of Oxford. Work at the University
of South Carolina was supported by NSF/EPSCoR under grant No. EPS-0296165 and USC NanoCenter.
\end{flushleft}

\end{document}